\documentclass[12pt]{iopart}
\usepackage{times}
\usepackage{graphicx}

\begin{document}

\title{Evidence for Majorana Neutrinos: Dawn of a new era in spacetime 
structure}\footnote{This written version combines Concluding Remarks 
as well Invited Talk presented at this conference.}

\author{ D.\ V.\ Ahluwalia}

\address{ 
Inter-University Centre for Astronomy and Astrophysics (IUCAA)\\
Post Bag 4, Ganeshkhind, Pune, India 411 007}

\begin{abstract}
We show that Majorana particles belong to the Wigner class
of fermions in which the charge conjugation and the parity operators
commute, rather than anticommute. Rigorously speaking,
Majorana spinors do not satisfy the Dirac equation [a result originally
due to M. Kirchbach, which we re-render here]. Instead, they
satisfy a different wave equation, which we derive. This allows us to
reconcile St\"uckelberg-Feynman interpretation with the
Majorana construct. We present several new properties of neutral
particle spinors and argue that discovery of Majorana particles
constitutes dawn of a new era in spacetime structure.
\end{abstract}


\def\beq{\begin{eqnarray}}
\def\eeq{\end{eqnarray}}


\def\s{\mbox{\boldmath$\displaystyle\mathbf{\sigma}$}}
\def\J{\mbox{\boldmath$\displaystyle\mathbf{J}$}}
\def\K{\mbox{\boldmath$\displaystyle\mathbf{K}$}}
\def\A{\mbox{\boldmath$\displaystyle\mathbf{A}$}}
\def\B{\mbox{\boldmath$\displaystyle\mathbf{B}$}}

\def\P{\mbox{\boldmath$\displaystyle\mathbf{P}$}}
\def\p{\mbox{\boldmath$\displaystyle\mathbf{p}$}}
\def\hp{\mbox{\boldmath$\displaystyle\mathbf{\widehat{\p}}$}}
\def\x{\mbox{\boldmath$\displaystyle\mathbf{x}$}}
\def\0{\mbox{\boldmath$\displaystyle\mathbf{0}$}}
\def\bv{\mbox{\boldmath$\displaystyle\mathbf{\varphi}$}}
\def\hbv{\mbox{\boldmath$\displaystyle\mathbf{\widehat\varphi}$}}

\def\bn{\mbox{\boldmath$\displaystyle\mathbf{\nabla}$}}

\def\bl{\mbox{\boldmath$\displaystyle\mathbf{\lambda}$}}
\def\bl{\mbox{\boldmath$\displaystyle\mathbf{\lambda}$}}
\def\br{\mbox{\boldmath$\displaystyle\mathbf{\rho}$}}
\def\1{1}
\def\bfhh{\mbox{\boldmath$\displaystyle\mathbf{(1/2,0)\oplus(0,1/2)}\,\,$}}

\def\mn{\mbox{\boldmath$\displaystyle\mathbf{\nu}$}}
\def\amn{\mbox{\boldmath$\displaystyle\mathbf{\overline{\nu}}$}}

\def\mne{\mbox{\boldmath$\displaystyle\mathbf{\nu_e}$}}
\def\amne{\mbox{\boldmath$\displaystyle\mathbf{\overline{\nu}_e}$}}
\def\rlh{\mbox{\boldmath$\displaystyle\mathbf{\rightleftharpoons}$}}

\def\wm{\mbox{\boldmath$\displaystyle\mathbf{W^-}$}}
\def\hh{\mbox{\boldmath$\displaystyle\mathbf{(1/2,1/2)}$}}
\def\h00h{\mbox{\boldmath$\displaystyle\mathbf{(1/2,0)\oplus(0,1/2)}$}}
\def\znbb{\mbox{\boldmath$\displaystyle\mathbf{0\nu \beta\beta}$}}



\section{Introduction}

The long sought-after signal from experiments on neutrinoless
double beta decay has finally been reported by the   Heidelberg-Moscow (HM)
collaboration  \cite{HM2001,HM2002}.
In its most natural explanation the HM events 
suggest
neutrinos to be fundamentally neutral particles in the sense of 
Majorana \cite{EM1937}. 

It is our intention to argue that the discovery a 
Majorana particle, taken to its
logical implications, opens a new era in the structure of spacetime.
It constitutes a discovery in which spacetime is not merely a 
classical object, a mere $SU(2)_R\otimes SU(2)_L$ realization 
of the Lorentz algebra (in the sense of Ryder \cite{LHR1996}), but
the underlying representation spaces exploit additional relative 
phases. These phases between the two $SU(2)$ building blocks encode in them
important C, P, and T properties. Furthermore, Majorana particles belong
to a new and unusual Wigner class -- a class necessary for implementing
supersymmetry. Even though neutrino itself may not be a supersymmetric 
particle, its Majorana nature tells us that spacetime does realize 
a construct that is central to construction of supersymmetric theories.

The Lorentz algebra, associated with the generators of rotation, 
$\J$, and boosts, $\K$, 
fails to incorporate fermionic fields. 
As is well known, see, e.g., Ref. \cite{LHR1996}, 
this circumstance is  remedied by the introduction of two $SU(2)$ 
generators:
\beq
&& SU(2)_R:\quad \A= \frac{1}{2}\left(\J+ i\K\right)\,, \\
&& SU(2)_L:\quad \B= \frac{1}{2}\left(\J- i\K\right) \,.
\eeq
The resulting algebra is no longer that of the Lorentz group. In the
notation of Ref.  \cite{LHR1996}, the Weyl spinors belong to $(1/2,0)$ and 
$(0,1/2)$ representation spaces; and the parity covariant spin-$1/2$
constructs belong to the $(1/2,0)\oplus(0,1/2)$ representation space.
The vector indexed objects, such as $x^\mu$, $A^\mu$, transform as 
$(1/2,1/2)$, and so on. Therefore, while the Lorentz algebra has served
us well, the underlying representation spaces for the quantum 
field theoretic description 
of nature belong, at least in the non-Planckian realm, to the two
$SU(2)$'s introduced above. In reference \cite{KA2002} we have taken
an {\em ab intio\/} look at the representation spaces which are appropriate
to the description of spin $1/2$ charged particles, vector particles, and
Rarita-Schwinger particles. Here,  we
present an {\em ab initio\/} formalism that embodies the 
original spirit of Majorana but extends/completes it in a non-trivial way. 
In doing so we build upon, but do not confine, to already 
existing original literature \cite{EM1937,Mc1957,Case1957} and exploit 
our experience in the spacetime structure of massive particles 
to benefit us \cite{KA2002,AK2001,DVA1996,AGJ1994,AJG1993}.

Based upon our studies in Refs. \cite{KA2002,AJG1993} we  
take as given that wave equation for spinors underlying the
description of spin one half  charged particles carries a symmetry under
the operation of $(1/2,0)\oplus(0,1/2)$-representation-space
charge conjugation operator:
\beq
{C} = 
\left(
\begin{array}{cc}
0_2 & i\,\Theta \\
-i\,\Theta & 0_2
\end{array}
\right) {K}\,.\label{cc}
\eeq
Here, operator $K$ complex conjugates any object that appears 
on its right, 
and  $\Theta$ is the Wigner's spin-$1/2$ time reversal 
operator\footnote{
For an arbitrary spin it is defined 
by the property $\Theta \J \Theta^{-1}= -\,\J^\ast$.
We refrain from  identifying $\Theta$ with ``$-\,i\, \sigma_2$,''
as is done implicitly in all considerations
on the subject -- see, e.g., Ref. \cite{PR1989} --  because
such  an identification does not exist for higher-spin $(j,0)\oplus(0,j)$ 
representation spaces. The existence of  Wigner time reversal
operator for all $j$, allows, for fermionic $j$'s,
the introduction of $(j,0)\oplus (0,j)$ neutral particles.}
\beq
\Theta=
\left(
\begin{array}{cc}
0 & -1 \\
1 & 0
\end{array}
\right)\,.\label{wt}
\eeq
We further assume that  the $C$ operator, and related $P$, and $T$ operators,
are intrinsic to  the $(1/2,0)\oplus(0,1/2)$ representation space up to certain
phases which may be fixed by additional physical requirements.
The mentioned phases may affect commutative and anticommutative 
properties of the $C$, $P$, and $T$ as
these properties depend not only on the form of the operators but also
on these phases and the  spinors on which the operators act upon. 

The operator  $C$ appeared on the physics scene not 
in expectation but as a surprise that lay hidden in spacetime 
symmetries and it revealed itself in the now famous Dirac construct
for spin one half. The well-known symmetry associated with
this operator brought into existence prediction of an entirely 
new type matter, called antimatter. Yet, the particles associated with
matter, and those associated with antimatter, are not eigenstates of
this very operator. Instead, this operator takes particle spinors into
antiparticle spinors and vice versa. The notion readily extends 
to a fully quantum field theoretic framework.

Beyond Dirac, St\"uckelberg in 1942 and Feynman in 1948 
proposed to interpret 
antiparticles as particles scattered backward in time
\cite{ECGS1942,RPF1948,RPF1949}. This latter proposal, as 
noted by Hatfield \cite{BH1992}, carries the advantage that it applies 
equally well to fermions as to bosons. However, already in 1937 
Majorana identified particle creation operators with antiparticle creation 
operators. In that quantum-field-theoretic proposal, Majorana did not 
consistently alter the  relevant representation space since he  
still used Dirac's 
$u_h(\p)$ and $v_h(\p)$ spinors. Moreover, 
given the time sequence of ideas, he could not
foresee the impact of  St\"uckelberg-Feynman interpretation 
for his proposal.
The former deficiency was remedied, though only partly, by 1957 papers 
of  McLennan and Case \cite{Mc1957,Case1957}, and by the 1996 work of 
Ref. \cite{DVA1996}. Here we hope to attend to all these question, and 
in the process bring to attention additional structure in the 
theory of neutral particles.

At this stage of the {\em paper\/}
it is, therefore, to be concluded that as far as neutral 
particles are concerned the existing state of theory 
is unsatisfactory. It calls for an {\em ab initio\/} construction
based on the eigenspinors  of the $C$ operator. After the reader has 
examined our presentation, it is our contention that 
she/he will find it absurd, though ``workable,''
to describe charged particles in terms of the neutral-particle 
framework we present. We carry the same sentiments for the existing 
description of neutral particles (for it carries a strong
dependence on charged-particle framework). 
Charged and neutral particles demand
their own independent frameworks. Once that is done one may, if one wishes,
seek differences and similarities between the two. But, not before. 
This {\em paper\/} undertakes this task.

\begin{table}
\begin{center}
\begin{tabular}{|c|c|} \hline
$(1/2,0)\oplus(0,1/2)$ &    $(1/2,0)\oplus(0,1/2)$ \\
$\{C,P\}=0$ &  $\left[C,P\right]=0$ \\  
{\sc Fermionic  Matter Fields} &  {\sc Fermionic Gauge Fields}\\ \hline
 $(1,0)\oplus(0,1)$ &  $(1/2,1/2)$ \\
$\{C,P\}=0$ & $\left[C,P\right]=0$ \\  
{\sc Bosonic Matter Fields} &  {\sc Bosonic Gauge Fields} \\ \hline
\end{tabular}
\end{center}
\label{tab1}
\caption{The diagonal Wigner blocks are the ordinary fermionic matter and 
bosonic  gauge fields, while the off-diagonal blocks refer to new
structure in spacetime (refer to text for details). 
This primitive block can be extended to 
incorporate higher-spin particles of supergravity.
 }
\end{table}

The stated assertion that  
discovery of a spin-1/2 neutral particle shall be a step into 
a new realm of spacetime structure, arises, in part, from our ability to 
construct Table \ref{tab1}. In that table
the diagonal Wigner blocks are the ordinary fermionic-matter and 
bosonic-gauge fields. 
The top off-diagonal block is one of the key results of this {\em paper.\/}
Supersymmetric fermionic gauge bosons live in that
block. The bottom off-diagonal block is populated by bosonic matter fields
and awaits experimental confirmation.
It was constructed in the  1993 paper cited as Ref. \cite{AJG1993}. 
The gauge aspect of the  top off-diagonal block
tentatively refers either to the gauginos, or for neutrinos
(should they be confirmed to belong there) it should be
interpreted as an internal fermionic line as it appears  
in neutrinoless double beta decay. 
The $C$ and $P$ operators belong to the indicated representation space for 
each of the Wigner blocks. Possibility, but without explicit construction 
(for which we take credit), of such blocks is due to Wigner, and his 
colleagues \cite{EPW1962}.

We now present a detailed and systematic development
of the theory of neutral particles.

\def\rb{\kappa^{\left(\frac{1}{2},0\right)}}
\def\lb{\kappa^{\left(0,\frac{1}{2}\right)}}

\section{Neutral particle spinors as Eigenspinors of charge conjugation
operator: $\bl{\boldmath (}\p{\boldmath )}$
and $\br{\boldmath (}\p{\boldmath )}$ }

The boost generator for the
$(1/2,0)$ representation space is $-i\,\s/2$, 
and  that for $(0,1/2)$  is  $+i\,\s/2$. Consequently, 
the respective boosts are
\beq
\kappa^{\left(\frac{1}{2},0\right)} &=&\exp\left(+\, 
\frac{\s}{2}\cdot\bv\right)= \sqrt{\frac{E+m}{2\,m}}
\left(\1_2+\frac{\s\cdot\p}{E+m}\right)
\,,\label{br}\\
\kappa^{\left(0,\frac{1}{2}\right)}&=&\exp\left(-\, 
\frac{\s}{2}\cdot\bv\right)= \sqrt{\frac{E+m}{2\,m}}
\left(\1_2-\frac{\s\cdot\p}{E+m}\right)\,,\label{bl}
\eeq
where the linear boost parameter is defined as:
\beq
\cosh(\varphi)=\frac{E}{m},\quad
\sinh(\varphi)=\frac{\vert\p\vert}{m},\quad 
{\hbv}
=\frac{\p}{\vert \p\vert}\,.\label{bp}
\eeq
The boosts take a particle at rest  to a particle
moving with momentum $\p$ in the ``boosted frame.'' 
We use the notation in which 
$\1_n$ and $0_n$ represent
$n\times n$ identity and null matrices, respectively. 

Thus each of the $SU(2)$'s is separately endowed with the dispersion
relation $E^2=\p^2+m^2$ as encoded in Eqs. (\ref{bp}) via the identity,
$\cosh^2\left(\varphi\right) - \sinh^2\left(\varphi\right) = 1$.

The explicit expressions 
for $\rb$ and $\lb$ allow for the observation: 
\beq
\left( \kappa^{\left(0,\frac{1}{2}\right)} \right)^{-1} = 
\left( \kappa^{\left(\frac{1}{2},0\right)} \right)^\dagger\,,\quad
\left( \kappa^{\left(\frac{1}{2},0\right)} \right)^{-1} = 
\left( \kappa^{\left(0,\frac{1}{2}\right)} \right)^\dagger \, .
\eeq
Further,   $\Theta$, the Wigner's spin-$1/2$ time reversal 
operator, has the property
\beq
\Theta \left[\s/2\right] \Theta^{-1} = -\, \left[\s/2\right]^\ast\,,  
\label{wigner}
\eeq
When combined, these observations imply that \cite{PR1989}: 

\begin{enumerate}

\item
If $\phi_L(\p)$ transforms as a left handed spinor, then
$\left(\zeta_\lambda \Theta\right) \,\phi_L^\ast(\p)$
transforms as a right handed spinor -- where, $\zeta_\lambda$ is
an unspecified phase.

\item
If $\phi_R(\p)$ transforms as a right handed spinor, then
$\left(\zeta_\rho \Theta\right)^\ast \,\phi_R^\ast(\p)$
transforms as a left handed spinor -- where, $\zeta_\rho$ is
an unspecified phase.
\end{enumerate}

As a consequence, the following spinors 
belong to the $(1/2,0)\oplus(0,1/2)$
representation space :
\begin{center}
\fbox{%
\begin{minipage}{385pt}
\vspace{-\abovedisplayskip}
\beq
\lambda(\p) =
\left(
\begin{array}{c}
\left(\zeta_\lambda \Theta\right) \,\phi_L^\ast(\p)\\
\phi_L(\p)
\end{array}
\right)\,,\quad
\rho(\p)=
\left(
\begin{array}{c}
\phi_R(\p)\\
\left(\zeta_\rho \Theta\right)^\ast \,\phi_R^\ast(\p)
\end{array}
\right)\,.
\eeq
\end{minipage}}
\end{center}
Demanding $\lambda(\p )$ and $\rho (\p )$ to 
be self/anti-self conjugate under  $C$, 
\beq
{C} \lambda(\p) = \pm  \lambda(\p)\,,\quad
 {C} \rho(\p) = \pm  \rho(\p)\,,
\eeq
restricts the phases, $\zeta_\lambda$ and
$\zeta_\rho$, to two values:
\begin{center}
\fbox{%
\begin{minipage}{385pt}
\vspace{-\abovedisplayskip}
\beq
\zeta_\lambda= \pm\,i\,,\quad \zeta_\rho=\pm\,i\,.
\eeq
\end{minipage}}
\end{center}
The plus sign in the above equation
yields self conjugate, $\lambda^S(\p)$ and $\rho^S(\p)$ 
spinors; while the minus
sign results in the anti-self conjugate spinors,  $\lambda^A(\p)$ and
$\rho^A(\p)$. 
\medskip

\noindent
Several remarks appear appropriate:

\begin{enumerate}

\item
The $\lambda^{S,A}(\p)$ and $\rho^{S,A}(\p)$ 
are eigenspinors of the charge 
conjugation operator, $C$, in the $(1/2,0)\oplus(0,1/2)$ representation space.
They are counterpart of the Dirac's $u(\p)$ and $v(\p)$ spinors, which are
eigenspinors of the charge operator  in the same 
representation space.

\item
The self-conjugate spinors are the standard textbook material.\footnote{
See, e.g., Eq. (1.4.52) of Ref. \cite{PR1989}.
However, we take issue with the colloquial assertion that ``Majorana 
spinors,'' i.e., $\lambda^S(\p)$, are Weyl spinors in four-component form.
Such a misunderstanding has perhaps arisen due to implicit, or
inadvertent, neglect of the  $\lambda^A(\p)$ spinors. 
A Weyl spinor transform as a $(1/2,0)$, or as a $(0,1/2)$, spinor; while
the neutral particle spinors transform as  $(1/2,0)\oplus(0,1/2)$ 
four-component spinors.
The Weyl space is a two dimensional representation space. Hence,
it cannot be spanned by four independent neutral particle spinors,
i.e., $ \lambda^S(\p)$ {\em and\/}  $\lambda^A(\p)$.
That honor belongs to the the four-dimensional $(1/2,0)\oplus(0,1/2)$ 
representation space.}
However,
they must be supplemented by anti-self conjugate spinors to span the entire
$(1/2,0)\oplus(0,1/2)$ representation space. Their neglect results in
internal inconsistencies and the wrong conclusion on the true number
of degrees of freedom for neutral particles. Any attempt to discard
the anti-self conjugate spinors would parallel a call to discard the 
$v(\p)$ spinors. The latter would amount to throwing away the antiparticles
from one's theory. This would not only go against the observed reality 
but would make the theory internally inconsistent. 
Similar conclusions shall be seen to hold for our theory, and we shall
duly examine the entire set of eigenspinors associated with $C$.
It shall, however, suffice to confine to the set, 
$\lambda^{S,A}(\p)$, or to the physically equivalent set, $\rho^{S,A}(\p)$.

\item 
The necessary presence of the Wigner time reversal operator
in the neutral particle spinors, as we shall sometime call the eigenspinors
of the  $C$ operator, endows them with their own unique time evolution.
We shall examine this aspect  below. 

\item
Both the the $\lambda^{S,A}(\p)$, as well as 
$u(\p)$ and $v(\p)$ spinors, can be expressed in any realization (i.e.,
in Weyl, in Dirac, or in Majorana realizations; 
or whatever realization serves a particular  task at hand). This brings in 
no new physics beyond convenience.

\end{enumerate}

\section{The explicit form of $\bl{\boldmath (}\p{\boldmath)}$ spinors}

To obtain explicit expressions for $\lambda (\p )$, we first write down 
the rest spinors. These are:
\beq
\lambda^S(\0) = 
\left(
\begin{array}{c}
+\,i \,\Theta \,\phi_L^\ast(\0)\\
\phi_L(\0)
\end{array}
\right)\,,\quad
\lambda^A(\0) = 
\left(
\begin{array}{c}
-\,i \,\Theta \,\phi_L^\ast(\0)\\
\phi_L(\0)\, 
\end{array}
\right)\, .
\eeq
Next, we choose the $\phi_L(\0)$ to be helicity eigenstates,
\beq
\s\cdot{\hp} \;\phi_L^\pm (\0)= \pm\;\phi_L^\pm(\0)\,,
\label{x}
\eeq  
and concurrently note that
\beq
\s\cdot\hp \, \Theta \left[\phi_L^\pm (\0)\right]^\ast
= \mp\, \Theta\left[\phi_L^\pm(\0)\right]^\ast\,.
\label{y}
\eeq 

\bigskip
\begin{quote}
{\em Derivation of Eq. (\ref{y}):\/}
Complex conjugating Eq. (\ref{x})
gives, 
\[
{\s}^\ast\cdot{{\hp}} \;\left[\phi_L^\pm (\0)\right]^\ast= 
\pm\;\left[\phi_L^\pm(\0)\right]^\ast\,.  
\]
Substituting for $\s^\ast$ from Eq. (\ref{wigner}) then results in,
\[
\Theta \s \Theta^{-1}\cdot{\hp} \,\left[\phi_L^\pm (\0)\right]^\ast
= 
\mp\,\left[\phi_L^\pm(\0)\right]^\ast\, .
\]
But $\Theta^{-1} = -\Theta$. So, 
\[
- \Theta \s \Theta
\cdot{\hp} \,\left[\phi_L^\pm (\0)\right]^\ast= 
\mp\,\left[\phi_L^\pm(\0)\right]^\ast \,.
\]
Or, equivalently,
\[
\Theta^{-1} \s \Theta
\cdot{\hp} \,\left[\phi_L^\pm (\0)\right]^\ast= 
\mp\,\left[\phi_L^\pm(\0)\right]^\ast \,.
\]
Finally, left multiplying  both sides of the preceding 
equation by $\Theta$, and moving $\Theta$ through 
${\hp}$, yields Eq. (\ref{y}).
\end{quote}

\bigskip
\noindent
That is, $\Theta \left[\phi_L^\pm (\0)\right]^\ast$
has opposite helicity of $\phi_L^\pm (\0)$.
Since $\s\cdot\hp$ commutes with the boost operator 
$\kappa^{\left(1/2,0\right)}$ the above result applies for all
momenta. In conjunction with the definition of the 
neutral spinors we are thus lead to the result that
neutral spinors are {\em not\/} single helicity objects. Instead,
they invite an interpretation of dual helicity spinors.
This shall allow for processes like neutrinoless double
beta decay.

In the process  we are led to four rest spinors. 
Two of which are self-conjugate, 
\beq
\lambda_{\{-,+\}}^S(\0) = 
\left(
\begin{array}{c}
+\,i \,\Theta \,\left[\phi^+_L(\0)\right]^\ast\\
\phi^+_L(\0)
\end{array}
\right)\,,\quad
\lambda_{\{+,-\}}^S(\0) = 
\left(
\begin{array}{c}
+\,i \,\Theta \,\left[\phi^-_L(\0)\right]^\ast\\
\phi^-_L(\0)\, 
\end{array}
\right)\, , \quad
\eeq
and the other two, which are anti-self conjugate,
\beq
\lambda_{\{-,+\}}^A(\0) = 
\left(
\begin{array}{c}
-\,i \,\Theta \,\left[\phi^+_L(\0)\right]^\ast\\
\phi^+_L(\0)
\end{array}
\right)\,,\quad
\lambda_{\{+,-\}}^A(\0) = 
\left(
\begin{array}{c}
-\,i \,\Theta \,\left[\phi^-_L(\0)\right]^\ast\\
\phi^-_L(\0)\, 
\end{array}
\right)\, .
\eeq
The first helicity entry refers to the $(1/2,0)$ transforming component of the
 $\lambda(\p)$, while the second entry encodes the helicity of
the $(0,1/2)$  component.

The boosted spinors are now obtained via the operation:
\beq
\lambda_{\{h,-h\}}(\p)=\left(
\begin{array}{cc}
\kappa^{\left(\frac{1}{2},0\right)} & 0_2 \\
0_2 & \kappa^{\left(0,\frac{1}{2}\right)}
\end{array}
\right)\lambda_{\{h,-h\}}(\0)\,.\label{z}
\eeq
In the boosts, we replace $\s\cdot\p$ by $\s\cdot{\hp}\,\vert \p\vert$,
and then exploit Eq. (\ref{y}). After simplification,
Eq. (\ref{z}) yields:
\begin{center}
\fbox{%
\begin{minipage}{385pt}
\vspace{-\abovedisplayskip}
\beq
\lambda_{\{-,+\}}^S(\p)
=
\sqrt{\frac{E+m}{2\,m}}\left(1-\frac{ \vert \p \vert}{E+m}\right)
\lambda_{\{-,+\}}^S(\0)\,,\label{lsup}
\eeq
\end{minipage}}
\end{center}
which, in the massless limit, {\em identically vanishes\/}, while
\begin{center}
\fbox{%
\begin{minipage}{385pt}
\vspace{-\abovedisplayskip}
\beq
\lambda_{\{+,-\}}^S(\p)
=
\sqrt{\frac{E+m}{2\,m}}\left(1+\frac{ \vert \p \vert}{E+m}\right)
\lambda_{\{+,-\}}^S(\0)\, ,
\label{lsdown}
\eeq
\end{minipage}}
\end{center}
does not.
We hasten to warn the reader that one should not be tempted to read the
two different prefactors to $\lambda^S(\0)$
in the above expressions as the boost operator that appears in Eq. 
(\ref{z}). For one thing, there is only one (not two) 
boost operator(s) in the
$(1/2,0)\oplus(0,1/2)$ representation space. The simplification that
appears here is due to a fine interplay between Eq. (\ref{y}), the boost
operator, and the structure of the $\lambda^S(\0)$.
Similarly, the anti-self conjugate set of the boosted spinors reads:
\begin{center}
\fbox{%
\begin{minipage}{385pt}
\vspace{-\abovedisplayskip}
\beq
\lambda_{\{-,+\}}^A(\p)
=
\sqrt{\frac{E+m}{2\,m}}\left(1-\frac{ \vert \p \vert}{E+m}\right)
\lambda_{\{-,+\}}^A(\0)\,,\label{laup}\\
\lambda_{\{+,-\}}^A(\p)
=
\sqrt{\frac{E+m}{2\,m}}\left(1+\frac{ \vert \p \vert}{E+m}\right)
\lambda_{\{+,-\}}^A(\0)\,.\label{ladown}
\eeq
\end{minipage}}
\end{center}
In the massless limit, the first of these spinors 
{\em identically vanishes\/}, while the second does not.
Representing the unit vector along $\p$, 
as, 
\beq
\hp =\Big(\sin(\theta)\cos(\phi),\,
\sin(\theta)\sin(\phi),\,\cos(\theta)\Big)\,,
\eeq
the $\phi^\pm_L(\0)$ take the explicit form:
\beq
\phi_L^+(\0) =
\sqrt{m} e^{i\vartheta_1} 
\left(
\begin{array}{c}
\cos(\theta/2) e^{-i\phi/2}\\
\sin(\theta/2) e^{i\phi/2}
\end{array}
\right)\,,\\
\phi_L^-(\0) =
\sqrt{m} e^{i\vartheta_2} 
\left(
\begin{array}{c}
\sin(\theta/2) e^{-i\phi/2}\\
-\cos(\theta/2) e^{i\phi/2}
\end{array}
\right)\,.
\eeq
On setting $\vartheta_1$ and $\vartheta_2$ to be zero --- a fact
that we explicitly note \cite{DVA1996} --- we find the following {\em 
bi-orthonormality\/} relations for the self-conjugate neutral 
spinors,
\begin{center}
\fbox{%
\begin{minipage}{385pt}
\vspace{-\abovedisplayskip}
\beq
 \overline{\lambda}^S_{\{-,+\}}(\p) \lambda^S_{\{-,+\}} (\p) = 0\,,
\quad 
\overline{\lambda}^S_{\{-,+\}}(\p) \lambda^S_{\{+,-\}} (\p) = + 2 i m
\,,&& \label{bo1}\\
 \overline{\lambda}^S_{\{+,-\}}(\p) \lambda^S_{\{-,+\}} (\p) = - 2 i m\,,
\quad 
\overline{\lambda}^S_{\{+,-\}}(\p) \lambda^S_{\{+,-\}} (\p) = 0
\,.\label{bo2}&&
\eeq
\end{minipage}}
\end{center}
Their counterpart for antiself-conjugate neutral spinors reads, 
\begin{center}
\fbox{%
\begin{minipage}{385pt}
\vspace{-\abovedisplayskip}
\beq
 \overline{\lambda}^A_{\{-,+\}}(\p) \lambda^A_{\{-,+\}} (\p) = 0\,,
\quad 
\overline{\lambda}^A_{\{-,+\}}(\p) \lambda^A_{\{+,-\}} (\p) = - 2 i m
\,,&&\label{bo3}\\
 \overline{\lambda}^A_{\{+,-\}}(\p) \lambda^A_{\{-,+\}} (\p) = + 2 i m\,,
\quad 
\overline{\lambda}^A_{\{+,-\}}(\p) \lambda^A_{\{+,-\}} (\p) = 0
\,,\label{bo4}&&
\eeq
\end{minipage}}
\end{center}
while all combinations  of the type 
$\overline{\lambda}^A(\p) \lambda^S(\p)$ and
$\overline{\lambda}^S(\p) \lambda^A(\p)$ identically vanish.

We take note that the bi-orthogonal norms of the Majorana spinors
are intrinsically {\em imaginary.}
The associated completeness relation is:
\def\da{{\{+,-\}}}
\def\ua{{\{-,+\}}}
\begin{center}
\fbox{%
\begin{minipage}{385pt}
\vspace{-\abovedisplayskip}
\beq
-\frac{1}{2 i m}&&
{\Bigg(}\left[\lambda^S_{\{-,+\}}(\p) \overline{\lambda}^S_\da(\p)
-\lambda^S_\da(\p) \overline{\lambda}^S_{\{-,+\}}(p)\right]   \nonumber\\ 
&& \quad -
\left[\lambda^A_{\{-,+\}}(\p) \overline{\lambda}^A_\da(\p)
-\lambda^A_\da (\p)\overline{\lambda}^A_{\{-,+\}}(\p)\right]{\Bigg)}
 = \1_4\,.\nonumber \\ \label{lc}
\eeq
\end{minipage}}
\end{center}

\section{Neutral particle spinors in Majorana realization}

The $\lambda^{S,A}(\p)$ obtained above are in Weyl realization
-- locally, in this section, we denote them by,  
$\lambda^{S,A}_{W}(\p)$. 
In  Majorana realization (subscripted by, $M$) these spinors are given by:
\beq
\lambda^{S,A}_{M}(\p) = {\mathcal S} \,\lambda^{S,A}_{W}(\p)\,,
\eeq
where
\beq
{\mathcal S} = \frac{1}{2} \left(
\begin{array}{cc}
\1_2 + i\Theta & \1_2 - i\Theta \\
-\left(\1_2 - i\Theta\right) & \1_2 + i\Theta 
\end{array}\right)\,.
\eeq 
The $\lambda^{S}_{M}(\p)$ are real, while $\lambda^{A}_{M}(\p)$
are pure imaginary. 

\section{The $\br{\boldmath (}\p{\boldmath )}$ spinors are not independent}

Now, $(1/2,0)\oplus(0,1/2)$ is a four dimensional representation space.
Therefore, there cannot be more than four independent spinors.
Consistent with this observation, we find that 
the $\rho(\p)$ spinors are related to the $\lambda(\p)$ spinors 
via the following identities:
\beq
&&\rho^S_\ua(\p) = - i \lambda^A_\da(\p)\,,\quad
\rho^S_\da(\p) = + i \lambda^A_\ua(\p),\label{id1}\\
&&\rho^A_\ua(\p) = + i \lambda^S_\da(\p)\,,\quad
\rho^A_\da(\p) = - i \lambda^S_\ua(\p)\,.\label{id2}
\eeq
Using these identities, one may immediately obtain the bi-orthonormality
and completeness relations for the $\rho(\p)$ spinors. 
In the massless limit, $\rho^S_\downarrow(\p)$ and $\rho^A_\downarrow(\p)$
{\em identically vanish.\/}
A particularly simple orthonormality, as opposed to bi-orthonormality,
relation exists between the 
$\lambda(\p)$ and $\rho(\p)$ spinors:
\beq
\overline{\lambda}^S_\ua(\p) 
\rho^A_\ua(\p) = -2 m = \overline{\lambda}^A_\ua(\p)
\rho^S_\ua (\p)\\  
\overline{\lambda}^S_\da(\p) 
\rho^A_\da(\p) = - 2 m = \overline{\lambda}^A_\da(\p)
\rho^S_\da (\p).
\eeq
An associated completeness relation also exists, and it reads:
\beq
-\frac{1}{2  m}&&
{\Bigg(}\left[\lambda^S_\ua(\p) \overline{\rho}^A_\ua(\p)
+\lambda^S_\da(\p) \overline{\rho}^A_\da(p)\right] \nonumber \\
&& \quad+
\left[\lambda^A_\ua(\p) \overline{\rho}^S_\ua(\p)
+\lambda^A_\da (\p)\overline{\rho}^S_\da(\p)\right]{\Bigg)} 
= \1_4\,.\nonumber\\
\label{lrcompleteness}
\eeq
The results of this section are in spirit of Refs.~
\cite{Mc1957,Case1957,AGJ1994,DVA1996}.

The completeness relation (\ref{lc}) confirms
that a physically complete theory of 
neutral particle spinors must incorporate the 
self as well as antiself conjugate spinors. However, one has a choice.
One may either work with the set $\{\lambda^S(\p),\lambda^A(\p\})$,
or with the physically and mathematically equivalent set,
  $\{\rho^S(\p),\rho^A(\p)\}$. One is also free to choose some 
appropriate combinations of neutral particle spinors 
from these two sets.

\section{Comparison with the Dirac framework: The $\bl{\boldmath (}\p
{\boldmath )}$ do
not satisfy Dirac equation}

{\em 
The main result of this section is a re-rendering of a proof given
my M. Kirchbach\cite{mkpc}. Any mistake, if any, that the reader may 
notice is entirely due to my failure.\/}

The bi-orthonormality relations (\ref{bo1}-\ref{bo4})
and the completeness relation (\ref{lc}) are counterpart
of the following relations for the charged, i.e. Dirac, particle spinors:
\beq
&& \overline{u}_h(\p)\, u_{h^\prime}(\p) =  +2 m
 \delta_{h {h^\prime}}\,,
\quad
\overline{v}_h(\p)\, v_{h^\prime}(\p) =  -2 m
 \delta_{h {h^\prime}}\,,\label{5}\\
&& \frac{1}{2 m}
\left[\sum_{h=\pm 1/2}
 u_{h}(\p) \overline{u}_h(\p) -
 \sum_{h=\pm 1/2} v_{h}(\p) \overline{v}_h(\p)\right] = \1_4\,.\label{1}
\eeq
Furthermore, if one wishes (with certain element of hazard to become 
apparent below), one can 
write the the {\em momentum-space} neutral spinor set
 $\{\lambda^S(\p),\lambda^A(\p\})$, in terms of charged particle spinor
{\em momentum-space} set $\{u(\p),v(\p)\}$. This task is best
accomplished by introducing the following -- to be used
only locally  -- notation:
\beq
&& d_1\equiv u_+(\p), \,
d_2\equiv u_-(\p), \,
d_3\equiv v_+(\p), \,
d_4\equiv v_-(\p)\,, \\
&& m_1\equiv \lambda^S_\ua(\p), \,
m_2\equiv \lambda^S_\da(\p), \,
m_3\equiv \lambda^A_\ua(\p), \,
m_4\equiv \lambda^A_\da(\p)\,.
\eeq
Then, the neutral particle spinors can be written as,
\beq
m_i= \sum_{j=1}^4 \Omega_{ij} d_j\,,\label{md}
\eeq
where

\begin{equation}
\Omega_{ij}=
 \cases{
+\left({1}/{2 m}\right) \overline{d}_j\, m_i \1_{4}\,, & \quad{\mbox for}
\,\, $j =1,2$\cr
-\left({1}/{2 m}\right) \overline{d}_j \,m_i \1_{4}\,, & \quad{\mbox for} 
\,\,$j =3,4$\cr}\,.
\end{equation}
In matrix form,  the $\Omega$ reads:
\beq
\Omega =
\frac{1}{2}\left(
\begin{array}{cccc}
\1_4 & -i_4 & -\1_4 & -i_4 \\
i_4 & \1_4 & i_4 & -\1_4 \\
\1_4 & i_4 & -\1_4 & i_4 \\
-i_4 & \1_4 & -i_4 & -\1_4 
\end{array}
\right)\,,\label{omega}
\eeq
where, $i_4 \equiv i \1_4$.
Equations (\ref{md}) and (\ref{omega}) immediately tell us that a
neutral particle momentum-space spinor is a linear combination
of the  charged particle momentum-space {\em particle and antiparticle\/} 
spinors. In momentum space, the charged-particle spinors 
are annihilated by $\left(\gamma^\mu p_\mu \pm m \1_4\right)$,
\beq
\cases{
\mbox{For particles:}
\quad\left(\gamma^\mu p_\mu - m \1_4\right) u(\p)=0\,,\quad\cr
\mbox{For antiparticles:}\quad
\left(\gamma^\mu p_\mu + m \1_4\right) v(\p)=0\,.\cr}\label{deqs}
\eeq
Since the mass terms carry opposite signs, 
hence are different for the particle and antiparticle, 
the neutral particle spinors cannot be annihilated by 
$\left(\gamma^\mu p_\mu - m \1_4\right)$, or, by
$\left(\gamma^\mu p_\mu + m \1_4\right)$. Moreover,
in the configuration space,
since the time evolution of the of $u(\p)$ occurs via
$\exp(-ip_\mu x^\mu)$ while that for   
$v(\p)$ spinors occurs via $\exp(+ip_\mu x^\mu)$ one cannot naively go from
momentum-space expression (\ref{md}) to its configuration space counterpart.
In fact several conceptual and technically subtle hazards 
are confronted if one begins to mix the two set of spinors. One ought to,
as we intend to and shall, develop the theory of neutral particle spinors
entirely in its own right. We thus end this digression by making part
of the above argument more explicitly. For that purpose we introduce:
\beq
M\equiv
\left(
\begin{array}{c}
m_1\\
m_2\\
m_3\\
m_4
\end{array}
\right)\,,\quad
D\equiv
\left(
\begin{array}{c}
d_1\\
d_2\\
d_3\\
d_4
\end{array}
\right)\,,
\Lambda\equiv
\left(
\begin{array}{cccc}
\gamma_\mu p^\mu & 0_4 & 0_4 & 0_4 \\
 0_4 & \gamma_\mu p^\mu & 0_4 & 0_4  \\
  0_4 & 0_4& \gamma_\mu p^\mu & 0_4   \\
  0_4& 0_4 & 0_4 & \gamma_\mu p^\mu   \\
\end{array}
\right)
\,.
\eeq
In this language, equation (\ref{md}) becomes
\beq
M=\Omega D\,.
\eeq
Now, applying from left the operator $\Lambda$ and using,
$\left[\Lambda,\Omega\right] =0$, we get
\beq
\Lambda M = \Omega \Lambda D\,.
\eeq
But, Eqs. (\ref{deqs}) imply
\beq
\Lambda D = 
\left(
\begin{array}{cccc}
 m\1_4 & 0_4 & 0_4 & 0_4 \\
 0_4 &  m\1_4 & 0_4 & 0_4  \\
  0_4 & 0_4& - m \1_4 & 0_4   \\
  0_4& 0_4 & 0_4 &    - m \1_4 \\
\end{array}
\right)\,.\label{above}
\eeq
Therefore, on using $D=\Omega^{-1} M$ we obtain,
\beq
\Lambda M = \Omega \left(\mbox{r.h.s. of Eq. \ref{above}}\right) \Omega^{-1} M
\,.
\eeq
An explicit evaluation of, 
$\mu\equiv \Omega \left(\mbox{r.h.s. of Eq. \ref{above}}\right) \Omega^{-1}$,
reveals it to be,
\beq
\mu= \left(
\begin{array}{cccc}
 0_4& -i m\1_4 & 0_4 & 0_4 \\
 i m \1_4 &  0_4 & 0_4 & 0_4  \\
  0_4 & 0_4& 0_4 & i m \1_4   \\
  0_4& 0_4 & -i m \1_4 &    0_4 \\
\end{array}
\right)\,.
\eeq
Thus, finally giving us the result,
\beq
\left(
\begin{array}{cccc}
\gamma_\mu p^\mu & 0_4 & 0_4 & 0_4 \\
 0_4 & \gamma_\mu p^\mu & 0_4 & 0_4  \\
  0_4 & 0_4& \gamma_\mu p^\mu & 0_4   \\
  0_4& 0_4 & 0_4 & \gamma_\mu p^\mu   \\
\end{array}
\right)
\left(
\begin{array}{c}
\lambda^S_\ua(\p) \\
\lambda^S_\da(\p) \\
\lambda^A_\ua(\p) \\
\lambda^A_\da(\p)
\end{array}
\right)
-
im \left(
\begin{array}{c}
- \,\lambda^S_\da(\p) \\
 \,\lambda^S_\ua(\p) \\
 \,\lambda^A_\da(\p) \\
- \,\lambda^A_\ua(\p)
\end{array}
\right) = 0\,,\nonumber \\
\eeq
which explicitly establishes the result that 
$\left(\gamma^\mu p_\mu \pm m \1_4\right)$ do not annihilate
the neutral particle spinors.\footnote{The result contained in the above 
equation confirms earlier result of Ref. \cite{VVD1995}.} 
The text-book assertions that Majorana mass term is 
`off-diagonal'' is a rough translation of this equation.

\section{Commutativity of C and P for neutral particle spinors}

The parity operation is slightly subtle for neutral particle spinors.

With a reminder to remarks made immediately after Eq. (\ref{wt}),
the parity operator in the $(1/2,0)\oplus(0,1/2)$ representation
space is,
\beq
P= e^{i\phi_P} \gamma^0 {\mathcal R}\,,
\eeq
where 
\beq
\gamma^0 = \left(\begin{array}{cc}
		 0_2 & \1_2 \\
		 \1_2 & 0_2
		 \end{array}
	    \right)\,.
\eeq
The ${\mathcal R}$ is defined as,
\beq
{\mathcal R} \equiv
\left\{ \theta\rightarrow\pi-\theta, \;\phi\rightarrow \phi+\pi,\;
p\rightarrow p\right\}\,.
\eeq
This has the consequence that eigenvalues, $h$,  of the helicity
operator
\beq
{\mathbf h} = \frac{\s}{2}\cdot\hp
\eeq
change sign under the operation of $\mathcal R$,
\beq
{\mathcal R}: h \rightarrow h^\prime = - h\,.
\eeq
Furthermore, 
\beq
P u_h(\p) = e^{i\phi_P} \gamma^0 {\mathcal R} u_h(\p) =
 e^{i\phi_P} \gamma^0 u_{-h}(-\p) = 
- i e^{i\phi_P} u_h(\p)
\eeq  
Similarly, 
\beq
P v_h(\p) = i e^{i\phi_P} v_h(\p)\,.
\eeq
We now require the eigenvalues of the $P$ to be real. This
fixes the phase factor,
\beq
e^{i\phi_P} = \pm i\,. \label{pf}
\eeq
The remaining ambiguity, as contained in the sign, 
still remains.  It is  fixed by recourse to text-book 
convention by taking the sign on the right-hand side
of Eq. (\ref{pf}) of to be positive. This very last choice shall
not affect our conclusions (as it should not). 
The parity operator is thus fixed to be,
\beq
P= i \gamma^0 {\mathcal R}\,.
\eeq
Thus, 
\beq
&& P u_h(\p) = +\, u_h(\p)\,,\label{peqa}\\
 && P v_h(\p) = -\, v_h(\p)\,.\label{peqb}
\eeq 
The consistency of  Eqs.  (\ref{peqa}) and (\ref{peqb}) requires,
\beq
\mbox{\sc Charged particle spinors}: \quad P^2= \,1_4\,,\quad
[\mbox{{\em cf.} Eq.(\ref{cf2})}]\,.\label{cf1}
\eeq 

To calculate the anticommutator, $\{C,P\}$, when acting on
the $u_h(\p)$ and $v_h(\p)$ we now need, in addition,
the action of $C$ on these spinors. This action
can be summarized as follows:
\beq
C :{\Bigg\{}\begin{array}{l}
  u_{+1/2}(\p) \rightarrow - v_{-1/2}(\p)\,,
 u_{-1/2}(\p) \rightarrow   v_{+1/2}(\p)\,,\\
  v_{+1/2}(\p) \rightarrow   u_{-1/2}(\p) \,,
 v_{-1/2}(\p) \rightarrow - u_{+1/2}(\p)\,.\label{ceq}
\end{array}
\eeq
Using Eqs. (\ref{peqa}), (\ref{peqb}),  
and (\ref{ceq}) one can readily obtain the 
action of anticommutator, $\{C,P\}$, on the four $u(\p)$ and 
$v(p)$ spinors. For each case it is found to vanish: $\{C,P\}=0$.

\bigskip\noindent
The P acting on the  neutral particle spinors
yields the result,
\begin{center}
\fbox{%
\begin{minipage}{385pt}
\vspace{-\abovedisplayskip}
\beq
P\lambda^{S}_\ua (\p)= +\, i\, \lambda^A_\da(\p)\,,
P\lambda^{S}_\da (\p)= -\, i \,\lambda^A_\ua(\p)\,,&&\label{peqla}\\
P\lambda^{A}_\ua (\p)= - \,i \,\lambda^S_\da(\p)\,,
P\lambda^{A}_\da (\p)= + \,i \,\lambda^S_\ua(\p)\,.\label{peqlb}
\eeq
\end{minipage}}
\end{center}
Following the same procedure as before, we now use  (\ref{peqla}),  
(\ref{peqlb}), and (\ref{ceq}) to 
evaluate the action of the commutator $[C,P]$ 
on each of the four neutral particle spinors.
We find it vanishes for each of them: $[C,P] =0$.
It confirms the claim we made in Table \ref{tab1}.

The commutativity and anticommutatitvity of the $C$ and $P$ operators
is a deeply profound result and it establishes that the theory of
neutral and charged particles must be developed in their own rights.
This is the task we have undertaken and are developing here in this
{\em Paper.\/}

\section{\label{pa}Parity asymmetry for neutral particle spinors}

Unlike the charged particle spinors,  Eqs.  (\ref{peqla}) and (\ref{peqlb}) 
reveal that neutral particle spinors are not eigenstates of $P$.   
Furthermore, a rather apparently  
paradoxical asymmetry is contained in these equations.
For instance, the second equation in (\ref{peqla}) reads:
\beq
P\lambda^{S}_\da (\p)= -\, i \,\lambda^A_\ua(\p)\,.
\eeq
Now, in a normalization-independent manner
\beq
\lambda^{S}_\da (\p) \propto 
\left(1+\frac{ \vert \p \vert}{E+m}\right)
\lambda_\da^S(\0)\,,
\eeq
while
\beq
\lambda^A_\ua(\p) \propto
\left(1-\frac{ \vert \p \vert}{E+m}\right)
\lambda_\da^A(\0)\,.
\eeq
Consequently, in the massless/high-energy limit the $P$-reflection of 
$\lambda^{S}_\da (\p)$ identically vanishes. 
The same happens to the $\lambda^{A}_\da (\p)$ spinors
under $P$-reflection. This situation is in sharp
contrast to the charged particle spinors.
The consistency of  Eqs.  (\ref{peqla}) and (\ref{peqlb}) requires 
$P^2 = - \1_4$ and in the process shows that the remaining two, i.e.
first and third equation in that set, do not contain additional
physical content:
\begin{center}
\fbox{%
\begin{minipage}{385pt}
\vspace{-\abovedisplayskip}
\beq
\mbox{\sc Neutral particle spinors}: \quad P^2=-\,1_4\,.
\quad
[\mbox{{\em cf.} Eq.(\ref{cf1})}]\,.\nonumber \\ \label{cf2}
\eeq
\end{minipage}}
\end{center}
That is, for neutral particle spinors:
\begin{center}
\fbox{%
\begin{minipage}{385pt}
\vspace{-\abovedisplayskip}
\beq
\mbox{\sc Neutral particle spinors}: \quad P^4=\,1_4\,,\,.\label{cf3}
\eeq
\end{minipage}}
\end{center}

The origin of the asymmetry under $P$-reflection 
resides in the fact that the $(1/2,0)\oplus(1/2,0)$
neutral particles spinors, in being dual helicity objects, 
combine Weyl spinors of {\em opposite\/}
helicities. However,  in the massless limit, the structures of 
$\kappa^{\left(\frac{1}{2},0\right)}$ and 
$\kappa^{\left(0,\frac{1}{2}\right)}$ 
force only positive helicity $\left({1}/{2},0\right)$-Weyl 
and negative helicity $\left(0,{1}/{2}\right)$-Weyl spinors
to be non-vanishing.
For this reason, in 
the massless limit the neutral particle spinors, $\lambda^S_\ua(\p)$
and  $\lambda^A_\ua(\p)$,
carrying negative helicity $\left({1}/{2},0\right)$-Weyl 
and positive helicity $\left(0,{1}/{2}\right)$-Weyl spinors
identically vanish.

So we have the following situation: The $(1/2,0)\oplus(0,1/2)$ 
is a $P$ covariant representation space.  Yet, in the  neutral
particle formalism, it carries
$P$-reflection asymmetry. 
This circumstance has a precedence in the Velo-Zwanziger observation, who
noted \cite{VZ1969},
``the main lesson to be drawn from our analysis is that special relativity
is not automatically satisfied by writing equations which transform 
covariantly.'' We conjecture that this asymmetry may underlie the
phenomenologically known parity violation.
Even though the latter is incorporated, by
hand, in the standard model of the electroweak interactions its true 
physical origin has remained unknown.

\section{A Master wave equation for spinors}

To study time evolution of neutral particle spinors we need 
appropriate wave equation. This we do in the following manner.
First, we obtain the momentum-space wave equation satisfied
by the $\lambda(\p)$ spinors. Next, we ascertain the time
evolution via a ``$p_\mu\rightarrow i   \partial_\mu$''
prescription.  

Since the method we use, and the results we obtain, appear somewhat
unusual we exercise extra care in presenting our results. 
We, therefore, present a unified method which applies not only 
to neutral particle spinors but applies equally well 
to other cases (such as the Dirac formalism). 
The method is a generalization of the textbook 
procedure \cite{LHR1996} with corrections noted in Refs. 
\cite{AJG1993,GGA1995,DVA1998,AK2001}.  

Thus, define a general $(1/2,0)\oplus(0,1/2)$ spinor
\beq
\chi(\p) = \left(
\begin{array}{c}
\chi^{\left(\frac{1}{2},0\right)}(\p)\\
 \chi^{\left(0,\frac{1}{2}\right)}(\p)
\end{array}
\right)\,,\label{eq1}
\eeq
such that  in particle's rest frame, where, $\p=\0$, by definition,
\beq
\chi^{\left(\frac{1}{2},0\right)}(\0)
={\mathcal  A}\;
 \chi^{\left(0,\frac{1}{2}\right)}(\0)\,. \label{eq2}
\eeq
Here, the $2\times 2$ matrix ${\mathcal A}$ 
encodes $C$, $P$, and $T$ properties of the spinor
and is left unspecified at the moment except that we require it to
be invertible.

Once 
$\chi^{\left(\frac{1}{2},0\right)}(\0)$ and $
\chi^{\left(0,\frac{1}{2}\right)}(\0)$ are specified
the $\chi^{\left(\frac{1}{2},0\right)}(\p)$ and $
\chi^{\left(0,\frac{1}{2}\right)}(\p)$ follow from,
\beq
\chi^{\left(\frac{1}{2},0\right)}(\p) =&&\rb\;
\chi^{\left(\frac{1}{2},0\right)}(\0) \,, \label{eq3}\\
\chi^{\left(0,\frac{1}{2}\right)}(\p) =&&\lb\;
\chi^{\left(0,\frac{1}{2}\right)}(\0)\,. \,,\label{eq4}
\eeq
Below, we shall need their inverted forms also. These we write
as follows:
\beq
&&\chi^{\left(\frac{1}{2},0\right)}(\0) =
\left(\rb\right)^{-1} \;\chi^{\left(\frac{1}{2},0\right)}(\p)\,,\label{eq3b}\\
&&  \chi^{\left(0,\frac{1}{2}\right)}(\0) =
\left(\lb\right)^{-1}\;\chi^{\left(0,\frac{1}{2}\right)}(\p)\,.\label{eq4b}
\eeq
Equation (\ref{eq2}) implies,
\beq
\chi^{\left(0,\frac{1}{2}\right)}(\0) ={\mathcal A}^{-1}
\chi^{\left(\frac{1}{2},0\right)}(\0)
\eeq
which on immediate use of (\ref{eq3b}) yields, 
\beq
\chi^{\left(0,\frac{1}{2}\right)}(\0) &=&
{\mathcal A}^{-1} \left(\rb\right)^{-1} 
\;\chi^{\left(\frac{1}{2},0\right)}(\p) \,.
\eeq
However, since 
\beq
\left(\rb\right)^{-1} = \lb
\eeq
we have:
\beq
\chi^{\left(0,\frac{1}{2}\right)}(\0) =
{\mathcal A}^{-1}\; \lb \;\chi^{\left(\frac{1}{2},0\right)}(\p)
\,.\label{eq79}
\eeq
Similarly,
\beq
\chi^{\left(\frac{1}{2},0\right)}(\0) =
{\mathcal A}\;\rb \;\chi^{\left(0,\frac{1}{2}\right)}(\p)
\,.\label{eq80}
\eeq
Substituting for  $\chi^{\left(\frac{1}{2},0\right)}(\0)$
from Eq. (\ref{eq80}) in
Eq. (\ref{eq3}) and re-arranging
gives:
\beq
-\;\chi^{\left(\frac{1}{2},0\right)}(\p) \;+ \;\rb {\mathcal A}\; \rb \;
\chi^{\left(0,\frac{1}{2}\right)}(\p) = 0\,;
\eeq
while similar use of Eq. 
(\ref{eq79}) in Eq. (\ref{eq4}) results in:
\beq
\lb\;{\mathcal A}^{-1} \;\lb \; \chi^{\left(\frac{1}{2},0\right)}(\p)
\;-\; \chi^{\left(0,\frac{1}{2}\right)}(\p)  =0\,.
\eeq
The last two equations when combined into a matrix form result in the
{\em momentum-space master equation for\/} $\chi(\p)$,

\begin{center}
\fbox{%
\begin{minipage}{385pt}
\vspace{-\abovedisplayskip}
\beq
\left(
\begin{array}{cc}
- \1_2 & \rb {\mathcal A}\; \rb \\
\lb\;{\mathcal A}^{-1} \;\lb &  -\1_2
\end{array}
\right)\chi(\p)=0\,.\label{meq}
\eeq
\end{minipage}}
\end{center}
Thus, the momentum-space equation for 
$\chi(\p)$ is entirely determined by the boosts
$\rb$ and $\lb$ and the CPT-property encoding matrix $\mathcal A$. 

\medskip
\noindent
We envisage the most general form of 
$
{\mathcal A}
$ 
to be a unitary matrix with determinant $\pm 1$:
\beq
{\mathcal A_\pm} = 
\left(
\begin{array}{cc}
a\, e^{i \phi_a} & \sqrt{\pm 1-a^2}\, e^{i\phi_b} \\
 - \sqrt{\pm 1-a^2}\, e^{- i\phi_b} & a\,  e^{- i \phi_a} 
\end{array}
\right)\,,\label{amat}
\eeq
with $a$, $\phi_a$, and $\phi_b$ real. The plus sign yields
Determinant of ${\mathcal A}$
to be $+1$, while the minus sign yields it to be $- 1$.
Inserting 
$
{\mathcal A}
$
from 
Eq. (\ref{amat}) into (\ref{meq}), 
we evaluate the determinant of the operator 
\beq
{\mathcal O}= \left(
\begin{array}{cc}
- \1_2 & \rb {\mathcal A}\; \rb \\
\lb\;{\mathcal A}^{-1} \;\lb &  -\1_2
\end{array}
\right)\,,
\eeq
and find it to be:
\beq
Det[{\mathcal O}]= \frac{
\left(
m^2+p^2-(2 m +E)^2
\right)^2\;
\left(m^2+p^2-E^2
\right)^2}{
\left({2 m (E + m)}\right)^4 }
\,.\nonumber\\
\eeq
The wave operator, 
${\mathcal O}$, supports two type of spinors. Those associated
with the usual Einsteinian dispersion relation, 
\beq
E^2=m^2+p^2\,, \quad \mbox{multiplicity =  4}
\eeq
and those
associated with:
\beq
E=\cases{-\;2 m -\sqrt{m^2+p^2}\,,\quad \mbox{multiplicity =  2}\cr
	 -\;2 m +\sqrt{m^2+p^2}\,,\quad \mbox{multiplicity  = 2}}\,.\label{nd}
\eeq
The origin of the new dispersion relation must certainly lie, or at least 
we suspect it to be so, in
the new $U(2)$ phases matrix. We shall see below that for the Dirac,
as well as Majorana, spinors only the Einsteinian dispersion relation
gets invoked. 

We hope to take up the other class of spinors, $\chi(\p)$, in a subsequent
study. Should something of physical interest emerge we shall report
it in an appropriate publication.

\section{Obtaining Dirac equation from  Master equation}

To give confidence to our reader in the physical content of the
Master equation we now apply it to the charged particle 
spinors of Dirac formalism. Once we do that we shall return
to the task of constructing momentum-space wave equation
for the $\lambda(\p)$.

The $\mathcal A$ can be read off from the Dirac rest spinors.
However, we remind the reader, that the writing down of the 
Dirac rest spinors, as shown by Weinberg
and also by our independent studies,
follows from the following two requirements:
\begin{enumerate}
\item[$\mathcal R_1$:]
The conservation of parity \cite{SW1995,DVA1998,AK2001}.
\item[$\mathcal R_2$:]
That,
in a quantum field theoretic framework, the Dirac field describe 
fermions \cite{SW1995}.
\end{enumerate}
These {\em physical requirements\/} determine   
$\mathcal A$ to be:
\beq
\vspace{-\abovedisplayskip}
{\mathcal A} =\cases{+ \;
\1_2\,,\quad\mbox{for}\,\,u(\p)\,\,\mbox{spinors} \cr
-\;\1_2\,,\quad\mbox{for}\,\,v(\p)\,\,\mbox{spinors}\,,}
\eeq
and correspond to ${\mathcal A}_+$ with $a=1$, $\phi_a=0$, and
$a=1$, $\phi_a=\pi$, respectively, with $\phi_b$ remaining arbitrary.
The subscript on $\mathcal A$ simply represents that its determinant is 
plus unity.

Using this information in the Master equation (\ref{meq}),
along with the explicit expressions for $\rb$ and $\lb$,
yields:
\beq
\left(
\begin{array}{cc}
-\;\1_2 & \exp\left(\s\cdot\bv\right)\\
\exp\left(-\;\s\cdot\bv\right) & -\;\1_2
\end{array}
\right)\,u(\p)=0\,,&&\label{ueq} \\
\left(
\begin{array}{cc}
\1_2 & \exp\left(\s\cdot\bv\right)\\
\exp\left(-\;\s\cdot\bv\right) & \1_2
\end{array}
\right)\,v(\p)=0\,.&&\label{veq}
\eeq
Exploiting the fact that $\s^2=\1_2$, and using the 
definition of the boost parameter $\bv$  given in Eqs. (\ref{bp}),
the exponentials that appear in the above equation take the form,
\beq
\exp\left(\pm \;\s\cdot\bv\right)
={\left(E\1_2 \pm \s\cdot\p\right)\over m}\,.\label{lin}
\eeq
Using these expansions in Eqs. (\ref{ueq}) and  
(\ref{veq}), multiplying both sides of the resulting equations
by $m$, using $p_\mu=\left(E,-\p\right)$, and
introducing:
\beq
\gamma^0 = \left(
\begin{array}{ll}
0_2 & \1_2 \\
\1_2 & 0_2
\end{array}
\right)\,,
\quad
\gamma^i = \left(
\begin{array}{cc}
0_2 & -\sigma_i \\
\sigma_i& 0_2
\end{array}
\right)\,,
\eeq
gives Eqs. (\ref{ueq}) and (\ref{ueq}) the form 
\beq
\left(p_\mu\gamma^\mu -m \1_4\right)\,u(\p)=0\,, && \\
\left(p_\mu\gamma^\mu +m \1_4\right)\,v(\p)=0\,.
\eeq
These are the well-known momentum space wave equations for 
the charged particle spinors (i.e. the Dirac equations). 
The {\em linearity\/} of these equations in $p_\mu$ is due to form of 
$\mathcal A$, and the property of Pauli matrices, $\s^2=\1_2$  -- see, Eq.
(\ref{lin}).

\section{Obtaining wave equation for neutral particle spinors from 
Master equation}

The requirement that the $\lambda(\p)$ be eigenstates of the 
charge conjugation operator completely determines  $\mathcal A$
for the neutral particle spinors to be:
\beq
{\mathcal A} = \zeta_\lambda\,\Theta\,\alpha\,,
\eeq
where 
\beq
\alpha= \left(
\begin{array}{cc}
\exp\left(i\phi\right) & 0 \\
0 & \exp\left(-\;i\phi\right)
\end{array}
\right)\,.
\eeq
Explicitly, 
\beq
{\mathcal A}^S_{-}  =
\left(
\begin{array}{cc}
0 & - i e^{-i \phi} \\
i e^{i\phi} & 0
\end{array}
\right)\,,\quad
{\mathcal A}^A_{-} =
\left(
\begin{array}{cc}
0 &  i e^{-i \phi} \\
- i e^{i\phi} & 0
\end{array}
\right)\,.
\eeq
The noted $\mathcal A$'s arise from the following choice of the
parameters $\{a,\,\phi_a,\,\phi_b\}$: $a=0,\, \phi_b= -\phi+\pi$ and
$a=0,\, \phi_b= -\phi$, respectively,  with $\phi_a$ remaining arbitrary.
The subscript on $\mathcal A$  is to remind that its determinant is 
minus unity.  
This difference -- summarized in Table 2 --
in $\mathcal A$, for Dirac and Majorana spinors, 
does not allow the 
$\lambda(\p)$ to satisfy the  Dirac equation.

\begin{table}
\begin{center}
\begin{tabular}{|c|c|c|c|c|} \hline\hline
{\sc Spinor type}  & $\mbox{Det}[{\mathcal A}]$  
& $a$ & $\phi_a$ & $\phi_b$ \\ \hline\hline
$\mbox{Dirac}$
 $u(\p)$ & $+1$ & $1$ & $0$      & $\mbox{arbitrary}$ \\\hline  
$\mbox{Dirac}$ $v(\p)$ & $+1$ &$1$ & $\pi$      & $\mbox{arbitrary}$ 
\\  \hline
$\mbox{Majorana}$
$\lambda^S(\p)$ & $-1$ & 0 & $\mbox{arbitrary}$ & $-\phi + 
\pi $\\\hline
$\mbox{Majorana}$ $\lambda^A(\p)$ & $-1$ & 0 & $\mbox{arbitrary}$ 
& $-\phi  $
\\\hline \hline
\end{tabular}
\end{center}
\label{tab2}
\caption{The parameters $\{a,\,\phi_a,\,\phi_b\}$. See text.
 }
\end{table}

Following the same procedure as above, and using
\beq
\exp\left(\pm\;\frac{\s\cdot\bv}{2}\right)
=
\frac{
\left(E+m\right)\1_2 
\pm \s\cdot\p}
{\sqrt{2m\left(E+m\right)}}\,,
\eeq
we obtain, instead:
\def\op{\left(p_\mu\gamma^\mu + m \gamma^0\right)}
\beq
\left[\op 
\widetilde{\mathcal A} \op - 2m \left(E+m\right)\1_4\right]\lambda(\p)=0
\,;\label{eqnn}
\eeq
where
\beq
\widetilde{\mathcal A}=\left(
\begin{array}{cc}
0_2 & {\mathcal A} \\
{\mathcal A}^{-1} & 0_2
\end{array}
\right)\,.\label{eqn}
\eeq
However, $\widetilde{\mathcal A}$ commutes with $\op$
\beq
\left[\op,\;\widetilde{\mathcal A}\right]=0\,.
\eeq
Therefore, Eq. (\ref{eqnn}) after due simplification becomes:
\begin{center}
\fbox{%
\begin{minipage}{385pt}
\vspace{-\abovedisplayskip}
\beq
\bigg[
\Big(p_\mu p^\mu + 2 m E + m^2\Big) 
\widetilde{\mathcal A} - 
 2m \left(E+m \right)\1_4\bigg]\lambda(\p)=0\,.\label{eqnb}
\eeq
\end{minipage}}
\end{center}
As a check, we calculate the determinant of the
operator acting on $\lambda(\p)$, 
and find
\beq
\mbox{Det}\bigg[
\Big(p_\mu p^\mu + 2 m E + m^2\Big) 
\widetilde{\mathcal A} - 
2m \left(E + m\right) \1_4\bigg]\nonumber \\
= 
\left(
m^2+p^2-(2 m +E)^2\right)^2\;\left(m^2+p^2-E^2\right)^2\,.\label{dr1}
\eeq
Furthermore, we  make the observation that for  all-four $\lambda(\p)$,
$
\widetilde{\mathcal A}\; \lambda(\p) = \lambda(\p).
$   
With this observation, Eq. (\ref{eqnb}) shows that each of
the four components of the
$\lambda(\p)$
satisfies the Klein-Gordon equation:
$
\left[\left(p_\mu p^\mu - m^2\right)\1_4\right] \lambda(\p) =0.
$ 
However, the latter equation should not be considered
the wave equation for the $\lambda(\p)$. The correct equation
is (\ref{eqnb}), and it is this equation when 
transformed to the configuration space which shall yield the
full time evolution.

\section{Insensitivity of Majorana spinors  to the direction
of time}

The plane waves for the $\lambda(\x,t)$ and $\rho(\x,t)$
are, $\exp(-i\, \epsilon\, p\cdot x)\lambda(\p)$ and $
\exp(-i\,\epsilon\, p\cdot  x)\rho(\p)$, where $\epsilon=\pm 1$ (depending 
upon whether the propagation is forward in time, or backward in time).
To determine $\epsilon$, it suffices to study the wave equations in the
rest frame of the particles. In that frame, the $\lambda(\x,t)$
satisfies the following (simplified)differential 
equation:\footnote{To obtain the simplified 
equation below we first multiplied 
the momentum-space wave equation by $2m (E+m)$. Then, we 
exploited the fact 
that in configuration space, $\p=\frac{1}{i} {\bn}$. 
When it acts  upon 
$\lambda(\0)$ the resulting eigenvalue 
vanishes (so we dropped this term), 
and that $E=i \frac{\partial}{\partial t}$.} 
\def\a{- 2 m \left(i \frac{\partial}{\partial t} + m\right)}

\def\b{- \zeta_\lambda e^{-i\phi}\left(m^2 - \frac{\partial^2}{\partial t^2}
+ 2 i m \frac{\partial}{\partial t}\right)}

\def\c{\zeta_\lambda e^{i\phi}\left(m^2 - \frac{\partial^2}{\partial t^2}
+ 2 i m \frac{\partial}{\partial t}\right)}

\def\oa{{\mathcal O}_a}
\def\ob{{\mathcal O}_b}
\def\oc{{\mathcal O}_c}
\beq
\left(
\begin{array}{cccc}
\oa & 0  & 0 & \ob  \\
0  & \oa & \oc & 0\\ 
0  & \ob & \oa & 0\\
\oc & 0 & 0 & \oa
\end{array}
\right)\lambda(\0) e^{-i \epsilon\, m t} = 0\label{te}
\eeq
where
\beq
&&\oa\equiv\a\,,\\
&&\ob\equiv\b\,,\\
&&\oc\equiv\c\,.
\eeq
This equation does not fix the sign of $\epsilon$.
It only determines $\epsilon^2$ to be unity. 
To convince the reader, we give an example result of a simple
calculation. Let's consider $\lambda(\0)$ to be $\lambda^S_\ua(\0)$.
Then, set $\zeta_\lambda = \zeta^S_\lambda=i$. For this example,
the time evolution Eq. (\ref{te}) gives:
\beq
m^2 \left(\epsilon^2 -1\right)\lambda^S_\ua(\0) e^{-i \epsilon m t} =0\,.
\eeq
This result does not determine sign of $\epsilon$.
That is, 
Majorana spinors are insensitive to the {\em forward\/} and
{\em backward\/} directions in time so important in the Feynman-St\"uckelberg
interpretation of particles and antiparticles.
The {\em conventional\/} 
distinction between particles and antiparticles
disappears.

\section{ 
Remarks for a quantum field theoretic description 
for neutral particles}

In the  Dirac theory of charged particles, the {\em positive\/}-definite norms
of the $u(\p)$ spinors and {\em negative\/}-definite norms of the
$v(\p)$ spinors {\em and \/} the 
anticommutativity of the annihilation and
creation operators plays a fundamental role in securing 
a theory with energy bounded from below. 
The Dirac-particle dual, $\overline{\psi}(\p)$ 
to a $(1/2,0)\oplus(0,1/2)$  spinor, $\psi(\p)$, is,
\beq
\overline{\psi}(\p)= \left[\psi(\p)\right]^\dagger \gamma^0.
\eeq 
It yields real-definite norm for the spinors inhabiting the
basis spinors of the Dirac's $(1/2,0)\oplus(0,1/2)$ representation space,
and it implies the stated spinorial properties.

In order to quantize the theory with neutral particle spinors we 
find it necessary (as has been verified through a detailed calculation)
to define {\em neutral-particle duals}  $\stackrel{\frown}{\lambda}(\p)$:
\beq
&&\stackrel{\frown}{\lambda}^S_\ua(\p) = +\; \overline\rho^A_\da(\p)\,,\quad
\stackrel{\frown}{\lambda}^S_\da(\p) = + \;\overline\rho^A_\ua(\p)\,,\\
&&\stackrel{\frown}{\lambda}^A_\ua(\p) = -\; \overline\rho^S_\da(\p)\,,\quad
\stackrel{\frown}{\lambda}^A_\da(\p) = -\; \overline\rho^S_\ua(\p)\,.
\eeq
The use of dispersion relation $E=\pm\sqrt{p^2+m^2}$, yields:
\beq
\begin{array}{cc}
\stackrel{\frown}{\lambda}^S_\ua(\p) {\lambda}^S_\ua(\p) 
=\stackrel{\frown}{\lambda}^S_\da(\p) {\lambda}^S_\da(\p) =+\;2 m\,,\\ 
\stackrel{\frown}{\lambda}^A_\ua(\p) {\lambda}^A_\ua(\p) 
=
\stackrel{\frown}{\lambda}^A_\da(\p) {\lambda}^A_\da(\p)=  -\;2 m\,.
\end{array}
\eeq
We construct the projectors:
\beq
&&\vspace{-1cm}P_S=\frac{1}{2 m} \left( 
		 {\lambda}^S_\ua(\p)  \stackrel{\frown}{\lambda}^S_\ua(\p)+
		 {\lambda}^S_\da(\p)  \stackrel{\frown}{\lambda}^S_\da(\p)
		\right){\Bigg\vert}_{E= \pm\sqrt{p^2+m^2}} \quad\\
&& P_A=\frac{-1}{2 m} \left( 
		 {\lambda}^A_\ua(\p)  \stackrel{\frown}{\lambda}^A_\ua(\p)+
		 {\lambda}^A_\da(\p)  \stackrel{\frown}{\lambda}^A_\da(\p)
		\right){\Bigg\vert}_{E=  \pm\sqrt{p^2+m^2}}\quad
\eeq
and verify that indeed, $P_S^2=P_S$ and $P_A^2=P_A$.  Furthermore,
these degrees of freedom form a complete set:
\beq
P_S+P_A= I_4\,.
\eeq

Elsewhere we shall report on the quantum field theory based on
these degrees of freedom \cite{ADKS}.

\section{Conclusion}

We showed that Majorana particles belong to the Wigner class
of fermions in which the charge conjugation and the parity operators
commute, rather than anticommute. We proved, that rigorously speaking,
Majorana spinors do not satisfy the Dirac equation. Instead, they
satisfy a different wave equation, which we derived. This allowed us to
reconcile St\"uckelberg-Feynman interpretation with the
Majorana construct. We presented several new properties of neutral
particle spinors and argued that  discovery of Majorana particles
constitutes dawn of a new era in spacetime structure.

\section*{Acknowledgments}

I have enjoyed numerous  in-person discussions with M. Kirchbach
on the subject of this manuscript. It is inevitable that this 
written version contains many of her insights and contributions,
without she being responsible for its content and presentation, in
any direct manner. For my failures my apologies to her. For her
time, patience, and insights, my deepest thanks to her.

I extend my warmly felt thanks to the local and international
organizers of the Beyond the Desert 2002. I also thank
and  congratulate with all the very best wishes, 
Hans Klapdor-Kleingrothaus, and his 
collaborators at the Heidelberg-Moscow collaboration, for presenting
us the positive signal  on neutrinoless double beta decay \cite{HM2001}, 
and for his
patience in explaining all the questions raised on the subject 
\cite{HM2002}. 

I extend my warmest thanks to Naresh Dadhich, and
to Parampreet Singh, for extended
discussions which led to a deeper understanding, to new insights, and
to new questions \cite{ADKS}.

 IUCAA's hospitality, 
where part of this work was done,
is recorded with appreciation. 
CONACyT (Mexico) is thanked for funding this research through Project
E-32076.

\end{document}